\documentclass[manuscript]{aastex}
\usepackage{epsfig}
\usepackage{amsmath}
\usepackage{color}
\begin{document}

\title{Towards the IR Detection of Carbonic Acid: Absorption and Emission Spectra}

\date{\today}
\author{Ryan C. Fortenberry}
\email{r410@olemiss.edu}
\affil{Department of Chemistry \& Biochemistry, University of
Mississippi, University, Mississippi 38677-1848, U.S.A.}
\author{Vincent J. Esposito}
\affil{NASA Ames Research Center, MS N245-3, Moffett Field, California 94035,
U.S.A.}

\begin{abstract}


With the recent radioastronomical detection of \emph{cis-trans-}carbonic acid (H$_2$CO$_3$) in a molecular cloud toward the galactic center, the more stable but currently unobserved \emph{cis-cis} conformer is shown here to have strong IR features. While the higher-energy \emph{cis-trans-}carbonic acid was detected at millimeter {\color{black}{and centimeter}} wavelengths, owing to its larger dipole moment, the vibrational structure of \emph{cis-cis-}carbonic acid is more amenable to its observation at micron wavelengths. Even so, both conformers have relatively large IR intensities, and some of these fall in regions not dominated by polycyclic aromatic hydrocarbons.  Water features may inhibit observation near the 2.75 $\mu$m hydride stretches, but other vibrational fundamentals and even overtones in the 5.5 $\mu$m to 6.0 $\mu$m range may be discernible with JWST data.  This work has employed high-level, accurately benchmarked quantum chemical anharmonic procedures to compute exceptionally accurate rotational spectroscopic data compared to experiment.  Such performance implies that the IR absorption and even cascade emission spectral features computed in this work should be accurate and will provide the needed reference for observation of either carbonic acid conformer in various astronomical environments.

\end{abstract}

\section{Introduction}

The recent detection \citep{Sanz23} of carbonic acid (H$_2$CO$_3$) towards
the galactic center in molecular cloud G+0.693-0.027 via the Yebes 40 m and
IRAM 30 m telescopes is the culmination of more than 25 years of speculation
about the presence of this molecule in the interstellar medium (ISM) beyond its
role in solar system ices \citep{Strazzulla96, Zheng07, Peeters2010, Jones14,
Sandford20}.

Carbonic acid is well-known to be significantly less stable than its related formic and other carboxylic acid molecular family members even though it is comprised of a ketone and two {\color{black}{hydroxyl groups}}.  The reason is that H$_2$CO$_3$ readily dissociates into water and carbon dioxide with a barrier of roughly 40 kcal/mol and thermodynamically lies about 10 kcal/mol above the H$_2$O and CO$_2$ products \citep{Kumar07}.  However, this implies that it should form relatively easily from moderately-processed ices comprised of these abundant and volatile molecules {\color{black}{\citep{Jones14, Wang21}}}.  Surface association of the $cis$- or $trans$-HOCO radical with the hydroxyl radical is the most likely formation pathway for carbonic acid \citep{Sanz23, Noble11, Ioppolo21}, but other mechanisms are not outside of the realm of possibility.  Regardless of the exact chemical mechanism, as most astrophysical ices are largely comprised of water mixed with some form of carbon oxide material (i.e.~methanol, formaldehyde, carbon monoxide, carbon dioxide, etc.), carbonic acid is almost certainly created in such ices.  Even though the vaporization of carbonic acid into the gas-phase could, again, quickly break this ketone-diol down into water and carbon dioxide as part of this process, some should survive long enough in the gas-phase to be observed. This recent detection is strong evidence to support the idea that it will persist in the gas-phase.

The conclusive observation of gas-phase, astronomical carbonic acid still
provides novel twists to the story of this seemingly simple but often
complicated molecule.  The most stable form of H$_2$CO$_3$ is actually the observed
\emph{cis-cis} conformer at roughly 2 kcal/mol lower in energy than the
\emph{cis-trans} arrangment \citep{Wallace21carbonic}, but the dipole moment of
\emph{cis-cis} is 0.2 D.  The \emph{cis-trans} form, on the other hand, has a
dipole moment of more than 3.0 D \citep{Mori11, Sanz23}.  Hence, the
higher-energy conformer has been observed. The \emph{trans-trans}
conformer is a transition state between equivalent \emph{cis-trans} conformers.
The relatively low-energy separation between the two minimum energy structures
implies that both will be present in most astronomical environments like the
ISM.  It follows then that if the higher-energy \emph{cis-trans} is observed,
the \emph{cis-cis} conformer will almost certainly be present, as well.
However, the small dipole moment implies that radioastronomical tools cannot be
utilized for its ready detection, but an upper-limit has been established
\citep{Sanz23}.  However, the advent of the \emph{James Webb Space Telescope}
(JWST) provides another avenue for potential observation of the lower energy
conformer.

While the rotational spectrum of \emph{cis-cis-}H$_2$CO$_3$ has been known for
more than a decade \citep{Mori11}, the IR spectrum of either conformer is
less-well established.  Computational \citep{Huber12, Reddy12, Sagiv18} and
Ne/Ar-matrix experimental \citep{Bernard11} studies have provided insights into
the vibrational features of carbonic acid, but gas-phase experimental or
high-level quantum chemical classification of the IR characteristics of any
conformer are currently lacking from the literature.  As a result, the best
chance to observe \emph{cis-cis}-carbonic acid will come from comparison of
JWST observations to newly reported spectral data.  The present work will
compute exceptionally high-accuracy quantum chemical characterization of both
conformers of carbonic acid in order to provide the means for the first {\color{black}{astronomical}} observation of \emph{cis-cis-}H$_2$CO$_3$ as well as IR observation of the
known \emph{cis-trans}.  Locating astronomical environments where both
conformers of carbonic acid reside as determined from remote observations
requiring comparison to accurate standards will also assist in analysis of how
UV photons affect its potential role in electronic spectral properties in ices
and disks \citep{Ioppolo21, Wallace21carbonic, Wallace22, Haney23}.

One of the most accurate quantum chemical approaches for producing
rovibrational spectral data utilizes coupled cluster theory at the singles,
doubles, and perturbative triples method with the explicitly correlated
formalism, CCSD(T)-F12b, \citep{Rag89, Shavitt09, ccreview, Adler07, Knizia09}
along with the core-correlating cc-pCVTZ-F12 basis set \citep{Dunning89,
Peterson2002, Yousaf08, Peterson08, Hill10}.  The triple$-\zeta$ basis set is
sufficient for large basis set approximations due to explicit correlation
accelerating convergence to the complete basis set limit \citep{Gyorffy18}.
These energies are then corrected for canonical CCSD(T) scalar relativity
\citep{Douglas74}.  These electronic energies are conjoined to a quartic force
field (QFF), a fourth-order Taylor series expansion of the potential portion of
the internuclear Watson Hamiltonian \citep{Fortenberry19QFF, Fortenberry22}.
This so-called F12-TcCR (for explicitly correlated, triple$-\zeta$, core
correlating, relativistic) QFF has been shown to provide exceptional accuracy
for comparison to known experimental standards often to within 1 cm$^{-1}$ or
better for organic molecules including the related methanediol where this
method confirmed its gas-phase synthesis \citep{Westbrook20, Watrous21, Zhu11,
Jian21, Davis22}.  Furthermore, IR emission can also now be computed to high
accuracy and done so in full cascade \citep{Mackie18Cascade} in order to
provide a clear comparison to JWST observations.  Both methods will be
employed in this work in order to give the clearest characterization yet as to
the IR observation of carbonic acid in both its lower-energy \emph{cis-cis} and
its radioastronomically-observed \emph{cis-trans} conformers.

\section{Computational Details}

\subsection{F12-TcCR QFF}

The F12-TcCR QFF procedure \citep{Watrous21} begins with a tight, core correlating
CCSD(T)-F12b/cc-pCVTZ-F12 geometry optimization of the molecular structure.
The geometry is then displaced by 0.005 \AA\ and 0.005 radians per the
following symmetry-internal coordinate system for $C_{2v}$
\emph{cis-cis-}H$_2$CO$_3$:
\begin{align}
S_1(a_1) &= \mathrm{O_1}=\mathrm{C}\\
S_2(a_1) &= \frac {1}{\sqrt{2}}[(\mathrm{C}-\mathrm{O_2})+(\mathrm{C}-\mathrm{O_3})]\\
S_3(a_1) &= \frac {1}{\sqrt{2}}[(\mathrm{O_2}-\mathrm{H_1})+(\mathrm{O_3}-\mathrm{H_2})]\\
S_4(a_1) &= \frac {1}{\sqrt{2}}[\angle(\mathrm{O_1}-\mathrm{C}-\mathrm{O_2})+\angle(\mathrm{O_1}-\mathrm{C}-\mathrm{O_3})] \\
S_5(a_1) &= \frac {1}{\sqrt{2}}[\angle(\mathrm{C}-\mathrm{O_2}-\mathrm{H_1})+\angle(\mathrm{C}-\mathrm{O_3}-\mathrm{H_2})] \\
S_6(b_2) &= \frac {1}{\sqrt{2}}[(\mathrm{C}-\mathrm{O_2})-(\mathrm{C}-\mathrm{O_3})]\\
S_7(b_2) &= \frac {1}{\sqrt{2}}[(\mathrm{O_2}-\mathrm{H_1})-(\mathrm{O_3}-\mathrm{H_2})]\\
S_8(b_2) &= \frac {1}{\sqrt{2}}[\angle(\mathrm{O_1}-\mathrm{C}-\mathrm{O_2})-\angle(\mathrm{O_1}-\mathrm{C}-\mathrm{O_3})] \\
S_9(b_2) &= \frac {1}{\sqrt{2}}[\angle(\mathrm{C}-\mathrm{O_2}-\mathrm{H_1})-\angle(\mathrm{C}-\mathrm{O_3}-\mathrm{H_2})] \\
S_{10}(b_1) &= OPB(\mathrm{O_1}-\mathrm{C}-\mathrm{O_2}-\mathrm{O_2}) \\
S_{11}(b_1) &= \frac {1}{\sqrt{2}}[\tau(\mathrm{O_1}-\mathrm{C}-\mathrm{O_2}-\mathrm{H_1}) - \tau(\mathrm{O_1}-\mathrm{C}-\mathrm{O_3}-\mathrm{H_2})] \\
S_{12}(a_2) &= \frac {1}{\sqrt{2}}[\tau(\mathrm{O_1}-\mathrm{C}-\mathrm{O_2}-\mathrm{H_1}) + \tau(\mathrm{O_1}-\mathrm{C}-\mathrm{O_3}-\mathrm{H_2})], 
\end{align}
where OPB stands for ``out-of-plane-bend'' and O$_1$ is understood to the be
the ketone oxygen.  This produces 4493 total points.  For $C_s$
\emph{cis-trans-}H$_2$CO$_3$ there are 8965 total points constructed from the
following simple-internal coordinate system:
\begin{align}
S_1(a') &= \mathrm{O_1}=\mathrm{C}\\
S_2(a') &= \mathrm{C}-\mathrm{O_2}\\
S_3(a') &= \mathrm{C}-\mathrm{O_3}\\
S_4(a') &= \mathrm{O_2}-\mathrm{H_1} \\
S_5(a') &= \mathrm{O_3}-\mathrm{H_2} \\
S_6(a') &= \angle(\mathrm{O_1}-\mathrm{C}-\mathrm{O_2}) \\
S_7(a') &= \angle(\mathrm{O_1}-\mathrm{C}-\mathrm{O_3}) \\
S_8(a') &= \angle(\mathrm{C}-\mathrm{O_2}-\mathrm{H_1}) \\
S_9(a') &= \angle(\mathrm{C}-\mathrm{O_3}-\mathrm{H_2}) \\
S_{10}(a'') &= OPB(\mathrm{O_1}-\mathrm{C}-\mathrm{O_2}-\mathrm{O_2}) \\
S_{11}(a'') &= \tau(\mathrm{O_1}-\mathrm{C}-\mathrm{O_2}-\mathrm{H_1}) \\
S_{12}(a'') &= \tau(\mathrm{O_1}-\mathrm{C}-\mathrm{O_3}-\mathrm{H_2}) 
\end{align}
with the understanding that H$_2$ is the hydrogen atom creating the $trans$
conformer.

At each point F12-TcCR energies are computed via MOLPRO2020.1
\citep{MOLPRO-WIREs, MOLPRO22}.  These are then fit via a least-squares
procedure, transformed into Cartesian coordinates, and treated with rotational
and vibrational perturbation theory at second-order, VPT2, \citep{Mills72,
Watson77, Papousek82} as available in the PBQFF program \citep{Westbrook23}
built upon the SPECTRO program \citep{spectro91}.  Fermi resonance polyads can
be treated within VPT2 in this program \citep{Martin98} for increased accuracy {\color{black}{and are detected automatically with refinements made by the 
user}}. For \emph{cis-cis-}carbonic acid, these include: 2$\nu_3$=$\nu_1$;
2$\nu_7$=$\nu_9$+$\nu_6$=$\nu_{11}$+$\nu_5$=$\nu_3$;
$\nu_9$+$\nu_7$=$\nu_{12}$+$\nu_8$=$\nu_4$;
2$\nu_{10}$=$\nu_{10}$+$\nu_8$=$\nu_5$;
$\nu_{12}$+$\nu_8$=$\nu_{12}$+$\nu_{10}$=$\nu_6$; \&
2$\nu_{11}$=2$\nu_{12}$=$\nu_7$.  For \emph{cis-trans-}carbonic acid, the Fermi
resonance polyads are comprised of:
2$\nu_7$=$\nu_9$+$\nu_4$=$\nu_9$+$\nu_5$=$\nu_{11}$+$\nu_4$=$\nu_{11}$+$\nu_6$=$\nu_3$;
2$\nu_8$=$\nu_9$+$\nu_7$=$\nu_{10}$+$\nu_8$=$\nu_{11}$+$\nu_7$=$\nu_{12}$+$\nu_8$=$\nu_4$;
2$\nu_{10}$=$\nu_{10}$+$\nu_8$=$\nu_{12}$+$\nu_8$=$\nu_5$;
2$\nu_{10}$=2$\nu_{12}$=$\nu_{10}$+$\nu_8$=$\nu_{12}$+$\nu_8$=$\nu_{12}$+$\nu_{10}$=$\nu_6$;
\& 2$\nu_{10}$=2$\nu_{11}$=2$\nu_{12}$=$\nu_{12}$+$\nu_{10}$=$\nu_7$. {\color{black}{Accuracy of these methods compared to experimental benchmarks
for F12-TcCR QFF VPT2 anharmonic vibrational frequencies are typically within
1.5\% of their experimental values \citep{Watrous21}.}}

\subsection{Anharmonic Cascade Emission Spectrum}


The simulation of an anharmonic cascade emission spectrum has been described
previously in detail by \cite{Basire11} and \cite{Mackie18}.  In this present
study, a new, composite scheme is used to produce the emission spectrum of
\emph{cis-cis} and \emph{cis-trans} carbonic acid.  This composite scheme
involves the use anharmonicity constants ($\chi$) and vibrational frequencies
from the F12-TcCR method and IR intensities from 2$^{nd}$ order M{\o}ller-Plesset
perturbation theory, MP2 \citep{MP2}, {\color{black}{which has been shown to
provide notable accuracy for less cost compared to higher-level methods for
intensities \citep{Finney16}}}.  The F12-TcCR frequencies are obtained
from the above-described QFF procedure.  The MP2 portion begins with
computation of the optimized geometry followed by computation of the harmonic
normal modes and, finally, that of the quadratic, cubic, and quartic normal
coordinate force constants giving the MP2/aug-cc-pVTZ QFF of \emph{cis-cis} and
\emph{cis-trans} carbonic acid utilizing Gaussian 16 \citep{g16, Dunning89,
cc-pVXZ, aug-cc-pVXZ}.  To properly treat resonances, a locally modified
version of SPECTRO \citep{spectro91} is used to compute the anharmonic
vibrational absorption spectrum. SPECTRO takes the optimized geometry and QFF
as input. The QFF is computed in normal mode coordinates, and a linear
transformation is performed to produce a Cartesian coordinate QFF. The
resonances are determined via two parameters: the difference in energy between
the two states ($\Delta$) set to the default value of 200 cm$^{-1}$, and the
minimum value of the interaction between the two states ($W$) is also set to
the default value of 10 cm$^{-1}$.  Resonance polyads have the added advantage
of allowing for intensity redistribution among modes {\color{black}{based on the contribution of each mode to the final perturbed mode frequency resulting from the resonance analysis}} to calculate the IR
intensities more accurately. This procedure produces the MP2/aug-cc-pVTZ
anharmonic IR intensities utilized in the cascade emission simulations.

A vibrationally excited molecule in a collision-free environment radiatively
relaxes by emitting IR photons until all of the energy is dissipated. The
probability of emitting an IR photon at a given internal energy is proportional
to the magnitude of the vibrational frequency of the corresponding normal mode
multiplied by the energy-dependent emission at the given internal energy. This
process is modeled via a straightforward cascade emission process
\citep{Cook98, Pech02, Basire11, Mackie18Cascade}.  In short, a Wang-Landau
walk is used to calculate the density of states followed by a second
Wang-Landau random walk to produce an energy-dependent spectrum.  From there,
an IR photon is chosen based on the criteria described above resulting in the
energy loss of the IR photon. At this new internal energy, the energy-dependent
emission spectrum is recalculated, giving a new set of emission probabilities.
This is repeated until the molecule has relaxed to its vibrational ground
state, and the entire process is repeated until the desired accuracy is
achieved yielding a cascade emission spectrum. Each simulation is run with
varying internal excitation energies of 1 eV, 2 eV, 3 eV, and 4 eV using $2 \times 10^6$
photons.

\section{Results}


%
%

\renewcommand{\baselinestretch}{1}
\begin{table}[ht]
\centering

\caption{The F12-TcCR, and Experimental Rotational Constants for
\emph{cis-cis-} and \emph{cis-trans-}Carbonic Acid.}
\label{rots}

\begin{tabular}{r c | c c | c c}
 & & \multicolumn{2}{c|}{\emph{cis-cis}} & \multicolumn{2}{c}{\emph{cis-trans}}\\
Constant & Units & F12-TcCR & Exp.$^a$ & F12-TcCR & Exp.$^b$ \\
\hline
$A_e$ & MHz & 12071.1 &            & 11877.8 & \\
$B_e$ & MHz & 11397.2 &            & 11493.4 & \\
$C_e$ & MHz & 5862.1  &            & 5841.2  & \\
$A_0$ & MHz & 11998.8 & 11997.0550 & 11780.4 & 11778.6808 \\
$B_0$ & MHz & 11311.5 & 11308.3803 & 11426.4 & 11423.1345 \\
$C_0$ & MHz & 5814.9  &  5813.828  & 5793.2  &  5792.0741 \\
\hline
$\Delta_{J}$ & kHz & 6.259 & 6.165 & 5.451 & 5.73  \\
$\Delta_{K}$ & kHz & 9.475 & 6.0   & 6.933 & 8.14  \\
$\Delta_{JK}$ & kHz&-3.533 &-1.5   & 0.578 &-1.16  \\
$\delta_{J}$ & kHz & 2.698 & 2.618 & 2.286 & 2.618 \\
$\delta_{K}$ & kHz & 5.153 & 6.28  & 5.643 & 6.28  \\
\hline
$\Phi_{J}$ & mHz  &   7.893 & &   3.156\\
$\Phi_{K}$ & mHz  & 181.354 & & 153.699\\
$\Phi_{JK}$ & mHz &  47.549 & &  66.346\\
$\Phi_{KJ}$ & mHz &-204.683 & &-195.106\\
$\phi_{j}$ & mHz  &   3.966 & &   1.605\\
$\phi_{jk}$ & mHz &  28.260 & &  33.132\\
$\phi_{k}$ & mHz  &  17.301 & &  20.206\\
\end{tabular}
\\$^a$\cite{Mori11}.
\\$^b$\cite{Mori09}.
\end{table}
\renewcommand{\baselinestretch}{2}


\renewcommand{\baselinestretch}{1}
\begingroup
\begin{table}

\caption{The F12-TcCR, Previous, and Experimental Fundamental Vibrational
Frequencies for \emph{cis-cis-}Carbonic Acid in cm$^{-1}$ with Intensities ($f$)
in km/mol.$^a$}

\label{ccfreqs}

\centering

\begin{tabular}{l c | c | c c | c | c}
 &             &          & \multicolumn{2}{c|}{MP2} & \\
 & Description & F12-TcCR & Freqs. & $f$ & $\omega$B97XD$^a$ & Exp.$^b$ \\
\hline
$\omega_{1}$ ($a_1$)  & O$-$H symm.~stretch      & 3826.2 & 3825.6 & 15  & 3833 & \\
$\omega_{2}$ ($b_2$)  & O$-$H antisymm.~stretch  & 3823.6 & 3825.2 & 208 & 3880 & \\
$\omega_{3}$ ($a_1$)  & C$=$O stretch            & 1846.1 & 1847.9 & 521 & 1860 & \\
$\omega_{4}$ ($b_2$)  & C$-$O antisymm.~stretch  & 1487.6 & 1474.0 & 144 & 1488 & \\
$\omega_{5}$ ($a_1$)  & C$-$O$-$H symm.~bend     & 1308.6 & 1297.4 & 31  & 1298 & \\
$\omega_{6}$ ($b_2$)  & C$-$O$-$H antisymm.~bend & 1182.2 & 1171.0 & 437 & 1177 & \\
$\omega_{7}$ ($a_1$)  & C$-$O symm.~stretch      & 989.2  & 992.5  & 20  & 1005 & \\
$\omega_{8}$ ($b_1$)  & CO$_3$ OPB               & 808.5  & 814.3  & 40  & 822  & \\
$\omega_{9}$ ($b_2$)  & O$-$C$-$O antisymm.~bend & 607.8  & 606.2  & 211 & 604  & \\
$\omega_{10}$ ($b_1$) & C$-$O$-$H symm.~OPB      & 603.0  & 619.2  & 47  & 612  & \\
$\omega_{11}$ ($a_1$) & O$-$C$-$O symm.~bend     & 552.4  & 556.5  & 6   & 556  & \\
$\omega_{12}$ ($a_2$) & H$-$O$-$C symm.~OPB      & 531.7  & 533.3  & --  & 539  & \\
$\nu_{1}$ ($a_1$)     & O$-$H symm.~stretch      & 3637.7 & 3639.5 & 13   & 3641 & 3634 \\
$\nu_{2}$ ($b_2$)     & O$-$H antisymm.~stretch  & 3634.9 & 3639.3 & 190  & 3638 & 3630 \\
$\nu_{3}$ ($a_1$)     & C$=$O stretch            & 1804.9 & 1812.6 & 401  & 1829 & 1783 \\
$\nu_{4}$ ($b_2$)     & C$-$O antisymm.~stretch  & 1438.5 & 1434.3 & 118  & 1452 & 1456 \\
$\nu_{5}$ ($a_1$)     & C$-$O$-$H symm.~bend     & 1261.2 & 1252.6 & 31   & 1234 & \\
$\nu_{6}$ ($b_2$)     & C$-$O$-$H antisymm.~bend & 1143.7 & 1130.6 & 403  & 1117 & 1187\\
$\nu_{7}$ ($a_1$)     & C$-$O symm.~stretch      & 963.0  & 968.1  & 15   & 987  & \\
$\nu_{8}$ ($b_1$)     & CO$_3$ OPB               & 793.5  & 799.0  & 36   & 802  & 798  \\
$\nu_{9}$ ($b_1$)     & O$-$C$-$O antisymm.~bend & 603.4  & 601.2  & 207  & 396  & \\
$\nu_{10}$ ($a_1$)    & C$-$O$-$H symm.~OPB      & 571.2  & 573.5  & 44   & 610  & \\
$\nu_{11}$ ($b_2$)    & O$-$C$-$O symm.~bend     & 546.8  & 548.8  & 6    & 547  & \\
$\nu_{12}$ ($a_2$)    & H$-$O$-$C symm.~OPB      & 509.8  & 512.5  & --   & 309  & \\
ZPVE                 &                          & 8664.6 &        &  &  & \\
$2\nu_7$             &                          & 1932.9 & 1933.4 & 10  &  & \\
$\nu_{11}+\nu_5$     &                          & 1820.1 & 1798.1 & 84  &  & \\
$\nu_{12}+\nu_8$     &                          & 1340.1 & 1307.4 & 45  &  & \\
$\nu_{11}+\nu_9$     &                          & 1151.4 & 1150.9 & 11  &  & \\
\hline

\end{tabular}
\\$^a$\cite{Huber12}.
\\$^b$Ne matrix data from \cite{Bernard11}.

\end{table}
\endgroup
\renewcommand{\baselinestretch}{2}

%
%

Table \ref{rots} immediately shows the quality of the F12-TcCR QFF results for
the known rotational constants.  Differences in the principle rotational
constants between quantum chemical computations and experiment are roughly 3 MHz or less or errors of
less than 0.03\%.  Hence, this quantum chemical approach is producing
exceptional accuracy boding well for other spectroscopic properties.  This
accuracy continues in large part for the quartic distortion constants in Table
\ref{rots}, and the sextic distortion constants are provided in this work for
the first time for high-precision comparison.  However, the IR, vibrational
features are much more novel and pressing for potential JWST observations.

\subsection{\emph{cis-cis-}Carbonic Acid Fundamental Vibrational Frequencies}

The F12-TcCR anharmonic frequencies and MP2 anharmonic intensities, however,
showcase that not only does carbonic acid have large IR intensities, but
\emph{cis-cis-}carbonic acid has larger intensities than the \emph{cis-trans}.
Hence, the lower-energy conformer should be more observable in the IR than the
radioastronomically-observed form.  Table \ref{ccfreqs} shows the vibrational
spectral data for \emph{cis-cis-}carbonic acid.  The $\nu_3$ C$=$O, ketone
stretch and the $\nu_6$ C$-$O$-$H antisymmetric bend exhibit the largest
intensities at roughly 400 km mol$^{-1}$.  For reference, the ``bright'' antisymmetric
stretch in water is a factor of more than five lower at roughly 70 km mol$^{-1}$.  As
such, the perceived accuracy of the F12-TcCR QFF implies that absorption
features at 1804.9 cm$^{-1}$ (5.54 $\mu$m) and 1143.7 cm$^{-1}$ (8.74 $\mu$m),
respectively for $\nu_3$ and $\nu_6$, should be markers of possible carbonic
acid.

The \emph{cis-cis-}carbonic acid conformer also has large intensities for many
other fundamental vibrational frequencies, including in the hydride stretching
region, especially for $\nu_2$ at 3634.9 cm$^{-1}$ (2.75 $\mu$m).
Unfortunately, this fundamental lies roughly 20 cm$^{-1}$ lower in frequency
than the less intense symmetric stretch of water at 2.73 $\mu$m.  Hence,
distinction between water and carbonic acid here would be challenging.  At the
other end of the frequency regime, the $\nu_9$ in-plane bending of the ketone
oxygen (the O$-$C$-$O antisymmetric bend) also exhibits a large intensity of
roughly 200 km mol$^{-1}$.  This 603.4 cm$^{-1}$ (16.57 $\mu$m) absorption bookends
the strong features for this conformer.  Other notable IR absorptions include
$\nu_4$ as well as the $\nu_{11}+\nu_5$ combination band.

Additionally, the correlation between the F12-TcCR QFF fundamental anharmonic
vibrational frequencies and those from Ne matrix data \citep{Bernard11} are
largely in line with what is accepted as differences brought about by matrix
shifts.  The biggest question mark comes from $\nu_6$ where the Ne matrix
frequency of 1187 cm$^{-1}$ is higher than all reported, computed frequencies.
This could be a misassignment, but the large intensity of the fundamental
transition should make this frequency stand out, casting doubt on any
improperly attributed peaks.  Hence, the relatively large matrix shift is likely real and may
result from the fundamentals changing frequency such that the
$\nu_{12}$+$\nu_8$=$\nu_{12}$+$\nu_{10}$=$\nu_6$ Fermi resonance polyad is
broken.  In any case, the correlation between levels of theory for F12-TcCR and
MP2 is also exceptional with all fundamentals in alignment of better than 8.0
cm$^{-1}$ save for, again, $\nu_6$ at 13.1 cm$^{-1}$.  Regardless, the QFF
appears to be performing well, implying that the rotational constants are reliable and that the present results will be valuable in the determination of IR
spectral features for this lowest energy form of carbonic acid.

\subsection{\emph{cis-trans-}Carbonic Acid Fundamental Vibrational Frequencies}

\renewcommand{\baselinestretch}{1}
\begingroup
\begin{table}

\caption{The F12-TcCR, Previous, and Experimental Fundamental Vibrational
Frequencies for \emph{cis-trans-}Carbonic Acid in cm$^{-1}$ with Intensities ($f$)
in km/mol.$^a$}

\label{ctfreqs}

\centering

\begin{tabular}{l c | c | c c | c | c}
 &             &          & \multicolumn{2}{c|}{MP2} & \\
 & Description & F12-TcCR & Freqs. & $f$ & $\omega$B97XD$^a$ & Exp.$^b$ \\
\hline
$\omega_{1}$ ($a'$)  & O$-$H$_2$ stretch  & 3825.1 & 3805.0 & 108 &  & \\
$\omega_{2}$ ($a'$)  & O$-$H$_1$ stretch  & 3820.1 & 3797.4 & 101 &  & \\
$\omega_{3}$ ($a'$)  & C$=$O stretch      & 1897.1 & 1880.6 & 474 &  & \\
$\omega_{4}$ ($a'$)  & C$-$O$-$H$_1$ bend & 1430.9 & 1408.9 & 305 &  & \\
$\omega_{5}$ ($a'$)  & C$-$O$-$H$_2$ bend & 1289.1 & 1270.2 & 119 &  & \\
$\omega_{6}$ ($a'$)  & C$-$O$_2$ stretch  & 1175.8 & 1159.4 & 201 &  & \\
$\omega_{7}$ ($a'$)  & C$-$O$_1$ stretch  & 977.2  & 967.9  & 46  &  & \\
$\omega_{8}$ ($a''$) & CO$_3$ OPB         & 795.1  & 790.2  & 24  &  & \\
$\omega_{9}$ ($a'$)  & O$-$C$-$O$_2$ bend & 613.4  & 605.8  & 9   &  & \\
$\omega_{10}$ ($a''$)& C$-$O$-$H$_1$ OPB  & 569.3  & 571.7  & 13  &  & \\
$\omega_{11}$ ($a'$) & O$-$C$-$O$_1$ bend & 549.7  & 544.6  & 27  &  & \\
$\omega_{12}$ ($a''$)& C$-$O$-$H$_2$ OPB  & 491.3  & 495.7  & 211 &  & \\
$\nu_{1}$ ($a'$)     & O$-$H$_2$ stretch  & 3637.7 & 3625.1 & 98  & 3770 & 3628 \\
$\nu_{2}$ ($a'$)     & O$-$H$_1$ stretch  & 3632.5 & 3615.4 & 88  & 3739 & \\
$\nu_{3}$ ($a'$)     & C$=$O stretch      & 1847.4 & 1845.1 & 242 & 1873 & 1836 \\
$\nu_{4}$ ($a'$)     & C$-$O$-$H$_1$ bend & 1385.7 & 1370.1 & 236 & 1386 & 1427 \\
$\nu_{5}$ ($a'$)     & C$-$O$-$H$_2$ bend & 1253.5 & 1220.8 & 176 & 1253 & \\
$\nu_{6}$ ($a'$)     & C$-$O$_2$ stretch  & 1134.6 & 1123.1 & 201 & 1123 & 1184 \\
$\nu_{7}$ ($a'$)     & C$-$O$_1$ stretch  & 948.1  & 943.1  & 45  & 969  & \\
$\nu_{8}$ ($a'$)     & CO$_3$ OPB         & 785.7  & 778.5  & 21  & 798  & 784  \\
$\nu_{9}$ ($a'$)     & O$-$C$-$O$_2$ bend & 608.2  & 600.7  & 8   & 614  & \\
$\nu_{10}$ ($a''$)   & C$-$O$-$H$_1$ OPB  & 545.8  & 540.8  & 24  & 632  & \\
$\nu_{11}$ ($a''$)   & O$-$C$-$O$_1$ bend & 544.8  & 539.0  & 26  & 540  & \\
$\nu_{12}$ ($a''$)   & C$-$O$-$H$_2$ OPB  & 477.2  & 474.4  & 195 & 529  & \\
ZPVE                 &                    & 8601.8 &        &  &  & \\
$2\nu_7$             &                    & 1911.4 & 1885.7 & 46  &  & \\
$\nu_{9}+\nu_5$      &                    & 1867.7 & 1829.7 & 162 &  & \\
$\nu_{10}+\nu_8$     &                    & 1336.3 & 1327.9 & 64  &  & \\
$\nu_{11}+\nu_{10}$  &                    & 1091.0 & 1140.0 & 12  &  & \\
$\nu_{12}+\nu_{11}$  &                    & 1022.8 & 1256.7 & 13  &  & \\
\hline

\end{tabular}
\\$^a$\cite{Huber12}.
\\$^b$Ne matrix data from \cite{Bernard11}.

\end{table}
\endgroup
\renewcommand{\baselinestretch}{2}

The story for \emph{cis-trans-}carbonic acid mirrors the \emph{cis-cis} quite
closely as shown in Table \ref{ctfreqs}.  The biggest difference comes in the
intensities of the fundamentals.  None are above 250 km/mol, but $\nu_3$,
$\nu_4$, $\nu_6$, and $\nu_{12}$ are all in the roughly 200 km mol$^{-1}$ to 250
km mol$^{-1}$ range.  The $\nu_3$ ketone stretch is, once more, a major contributor to
the overall intensity budget for this IR spectrum.  This
\emph{cis-trans-}carbonic acid C$=$O stretch is black-shifted to 1847.4
cm$^{-1}$ (5.41 $\mu$m) compared to the other conformer, but these distinct
differences are still not too far away from one another such that this
could make for a nice pairing that would aid in gas-phase observation of both
conformers at once.  The $\nu_9+\nu_5$ combination band at 1867.7 cm$^{-1}$
(5.35 $\mu$m) will also add features to the 5 $\mu$m region with its intensity
of 162 km mol$^{-1}$, again more than twice the standard candle in the antisymmetric stretch of water.

The breaking of the symmetry from $C_{2v}$ in the \emph{cis-cis} to $C_s$ here
in the \emph{cis-trans} decouples modes that would contribute to larger
intensities, but similar motions are producing the largest intensity
transitions.  For instance, the $\nu_4$ and $\nu_5$ C$-$O$-$H bends have
similar intensities but are split 140 cm$^{-1}$ apart.  Differently, the
$\nu_{12}$ OPB motion is the only one of its kind to showcase a large
intensity.  This fundamental at 477.2 cm$^{-1}$ (20.96 $\mu$m) is also the
lowest-frequency for either form of carbonic acid.

Correlation between F12-TcCR and MP2 QFFs is similar in this conformer as it is
in \emph{cis-cis-}carbonic acid.  Comparison to Ne matrix data still exhibits a
large matrix shift for $\nu_6$, but this conformer also has a large matrix
shift in $\nu_4$, as well.  These matrix shifts are all to lower-frequency gas-phase
fundamentals implying that either the polyads shift or the matrix is behaving
in non-standard ways; {\color{black}{further gas-phase experimental characterization for IR fundamental frequencies would help to settle these discrepancies}}.  Regardless, the computed and experimental frequencies
are within a few dozen cm$^{-1}$ of one another, and the QFF is not exhibiting
any questionable behaviors.  As such, the presently-computed frequencies should
be reliable proxies for gas-phase results potentially more so than the matrix
data.  The theoretical results are also complete and step beyond the density
functional theory computations utilized to assign the matrix spectral features
\citep{Bernard11}.

\subsection{Cascade Emission Spectra}

\begin{figure}
\centering
\caption{Total (top) and 2000 - 300 cm$^{-1}$ range (bottom) anharmonic cascade emission spectra of \emph{cis-cis-}carbonic
acid at excitation energies of 1, 2, 3, and 4 eV calculated with the F12-TcCR
\& MP2/aug-cc-pVTZ composite scheme. The intensity of the spectra from 7000 -
3600 cm$^{-1}$ are expanded by a factor of 100 to show the detail of the minor
features.}
\includegraphics[width=4.5in]{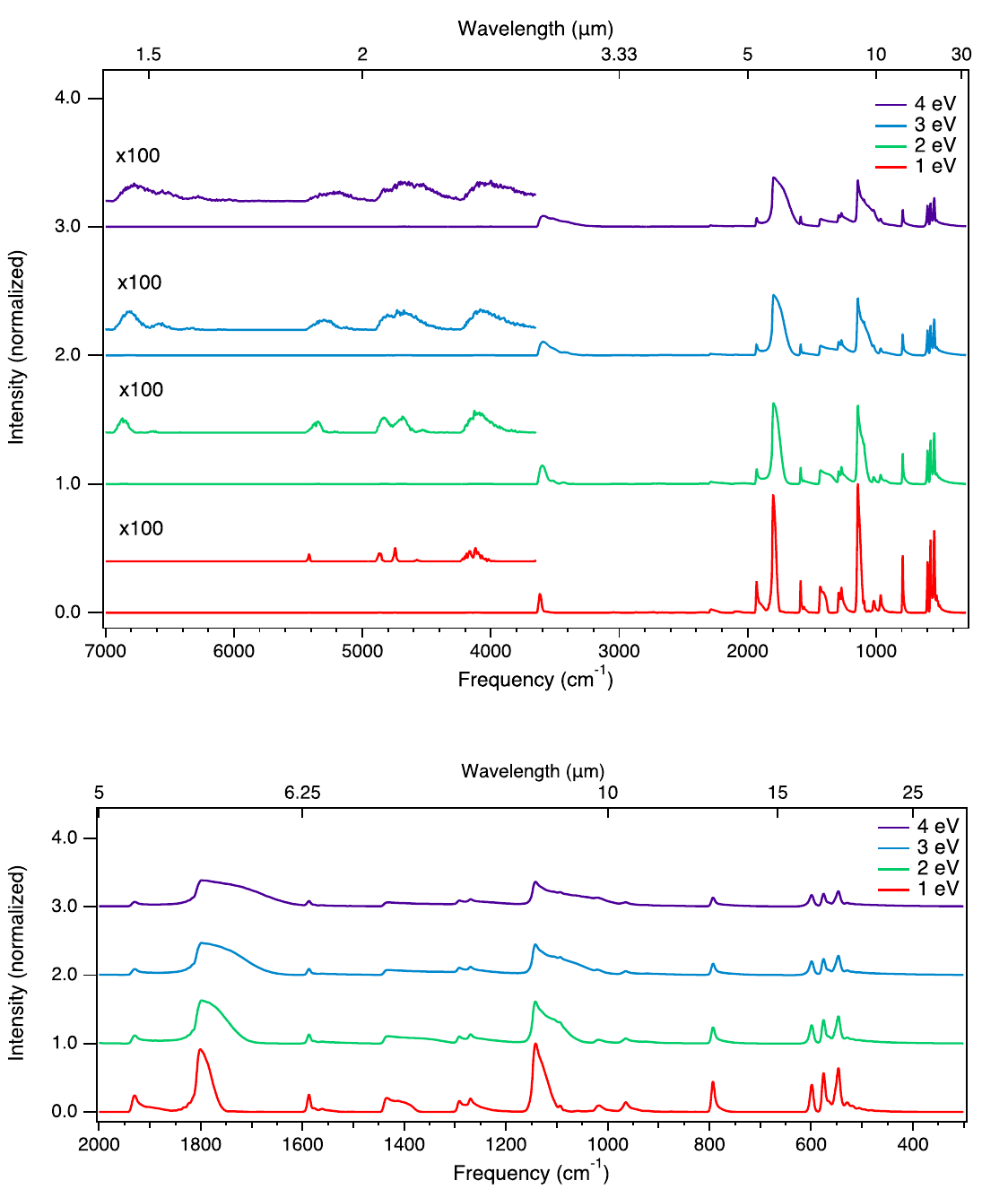}
\label{ccemission}
\end{figure}


The full emission spectrum of \emph{cis-cis-}carbonic acid is given in the tope of Figure
\ref{ccemission} showcasing how the above-discussed F12-TcCR anharmonic
fundamental vibrational features will emit when excited to the given internal energy.  The tallest emission features line up with the initial energy levels
discussed above (and given in Table \ref{ccfreqs}), but the full cascade
computed herein gives shape to the emissions.  Of course, the most notable
features correspond to $\nu_3$ and $\nu_6$, and as more energy is added to the
initial excitation, the red wings of the emission are broader as shown previously \citep{Mackie22}.  They even begin
to mix and combine beyond 2 eV.  Additionally, the low-frequency range from 600
cm$^{-1}$ and below (16.67 $\mu$m and longer) shows intensity borrowing from
$\nu_9$ given to $\nu_{10}$ and $\nu_{11}$ such that the least intense
absorption in $\nu_{11}$ (6 km mol$^{-1}$) actually has the most intense emission for
this triplet of emission peaks shown in more detail in the bottom of Figure \ref{ccemission}.  

\begin{figure}
\centering
\caption{Total (top) and 2000 - 300 cm$^{-1}$ range (bottom) anharmonic cascade emission spectra of \emph{cis-trans-}carbonic
acid at excitation energies of 1, 2, 3, and 4 eV calculated with the F12-TcCR
\& MP2/aug-cc-pVTZ composite scheme. The intensity of the spectra from 7000 -
3600 cm$^{-1}$ are expanded by a factor of 100 to show the detail of the minor
features.}
\includegraphics[width=4.5in]{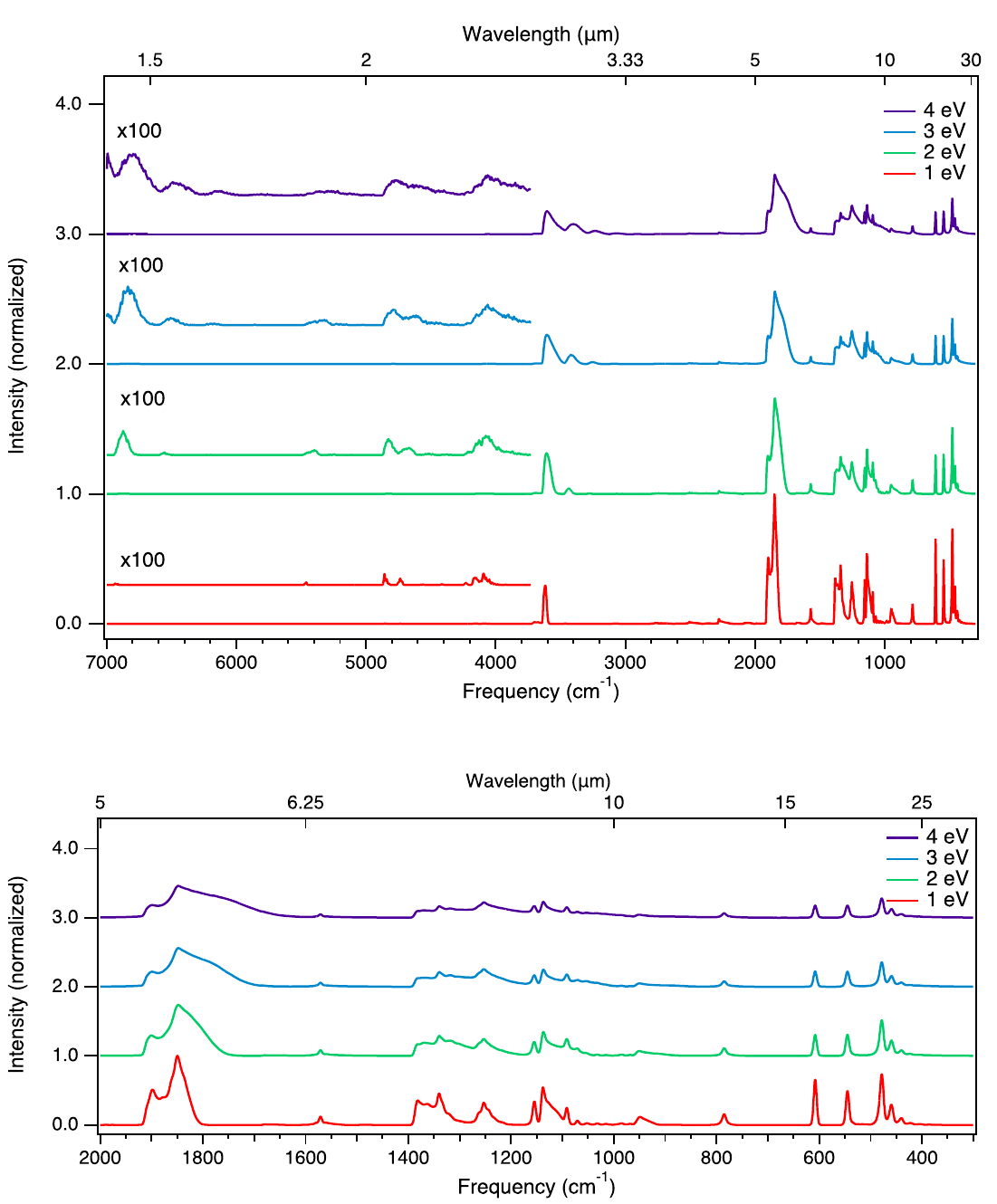}
\label{ctemission}
\end{figure}


Figure \ref{ctemission} showcases the same data for the \emph{cis-trans}
conformer.  While similar in many ways, the slight shifts of the fundamentals
between molecular structures are born out in the emissions.  Here, the leading,
tall peak at $\sim$1900 cm$^{-1}$ ($\sim$5.5 $\mu$m) is mixed with the
preceding and trailing ones since $2\nu_7$, $\nu_9+\nu_6$, and the fundamental
$\nu_3$ are all emitting within the same region.  As a consequence of this and
also as the internal energy increases, the width of the red wing
increases notably due to the presence of all three of these modes where the bottom of Figure
\ref{ctemission} highlights this more clearly.  Additionally, the hydride
stretching region at 3600 cm$^{-1}$ exhibits more structure in this less stable conformer
 than the \emph{cis-cis}.  While a quick visual
inspection could lead one to think that Figures \ref{ccemission} and
\ref{ctemission} are qualitatively similar, they have some differences on the
more quantitative scale that will show distinct vibrational structure upon
observation.

These emission spectra allow for direct comparison to JWST (or archival,
predecessor mission) observations as well as laboratory simulations for
interstellar spectral data.  The most obvious difference between these carbonic
acid emissions and those from IR observations is that the hydride stretches are
at shorter wavelengths due to the nature of O$-$H stretching as opposed to
C$-$H stretching at $\sim3.3$ $\mu$m believed to originate with polycyclic
aromatic hydrocarbons (PAHs) \citep{Allamandola99, Peeters04, Tielens08,
Boersma11, Allamandola21}.  Hence, while the hydride stretches in carbonic acid
are not the most intense or the brightest emitters for these molecules, they
may be among the best ones to observe due to a combination of the higher
resolution of JWST's NIRSpec instrument and the lack of crowding by most PAH
features.  Water will, again, affect this region of the spectrum, but the
resolution of JWST may allow for these to be distinguished.  Unfortunately, for
either isomer of carbonic acid, interstellar IR emissions from 5.5 $\mu$m to
6.0 $\mu$m are populated with what are believed to be some PAH features, but
these $\sim5.5$ $\mu$m features are not as numerous or strong as those at 3.3
$\mu$m, 6.2 $\mu$m, 7.7 $\mu$m, 8.6 $\mu$m, etc.  Hence, carbonic acid may yet
be observable with JWST's MIRI instrument in this range for both conformers'
strongest emission features.  The sub 1400 cm$^{-1}$ (7.1 $\mu$m and longer)
features of either carbonic acid isomer will likely be completely covered by
PAH features but could play a role in these regions, as well.

\section{Conclusions}

The IR wavelengths around 2.75 $\mu$m or between 5.5 $\mu$m and 6.0 $\mu$m
would likely be the best region to observe emission from carbonic acid in
either conformer.  The lower-energy, as-of-yet unobserved \emph{cis-cis}
conformer has larger IR intensities than the observed, less stable
\emph{cis-trans} conformer implying that either JWST NIRSpec or MIRI
observations could indicate the presence of \emph{cis-cis-}H$_2$CO$_3$, whereas
radioastronomical observations likely cannot.  While both molecules exhibit
similar features qualitatively, they have little direct overlap quantitatively.
Most fundamental vibrational transitions differ between the molecules by 50
cm$^{-1}$ or so making spectral deviations between conformers readily discernible with JWST.  While most of the features for carbonic acid
are in regions dominated by PAHs, the higher-frequency, shorter-wavelengths,
again, are the best bet for observing this molecule in the IR provided that
water features do not cause different opacity issues for carbonic acid.

The presence of \emph{cis-cis-}carbonic acid is almost a given with the known
observation of the \emph{cis-trans} conformer.  The barrier to rotation is
fairly low, and similar chemical processes should produce both conformers.
While the formation of carbonic acid from \emph{trans-}HOCO and hydroxyl
radicals on a grain surface would lead preferentially to the
\emph{cis-trans-}carbonic acid conformer, numerous other reactions can produce
this molecule which tautomerizes to astrochemically abundant carbon dioxide and
water.  Carbonic acid in any form stands to greatly enhance astrobiological
insights and expand the census of relatively simply organic molecules in space.
Its possession of a ketone as well as two alcohols makes it a useful
intermediate between carbon dioxide/water and prebiotic species, and this
molecule is likely present in astronomical environments stretching from disks
to planetary atmospheres to, obviously, molecular clouds in the ISM.  Now that
the full IR spectrum in absorption and emission has been produced herein,
further exploration for its presence in space can be undertaken.

\section{Acknowledgements}

RCF acknowledges support from NASA grants NNX17AH15G \& NNH22ZHA004C, NSF grant
OIA-1757220, and start-up funds provided by the University of Mississippi.  VJE
acknowledges support from the NASA Postdoctoral Program.  Both authors,
RCF and VJE, acknowledge support from the Internal Scientist Funding Model (ISFM) Laboratory Astrophysics Directed Work Package at NASA Ames (22-A22ISFM-0009).
Additionally, the authors would like to thank Izaskun M. Jim\'{e}nez-Serra of El Centro de Astrobiolog\'{i}a (CAB), INTA-CSIC in Spain for inspiration for and discussions of this research project as well as Alexandros Maragkoudakis of the NASA Ames Research Center for ideas about how these data could influence JWST observations.

\bibliographystyle{apj}
\bibliography{refs}

\begin{thebibliography}{62}
\expandafter\ifx\csname natexlab\endcsname\relax\def\natexlab#1{#1}\fi

\bibitem[{Adler {et~al.}(2007)Adler, Knizia, \& Werner}]{Adler07}
Adler, T.~B., Knizia, G., \& Werner, H.-J. 2007, J. Chem. Phys., 127, 221106

\bibitem[{Allamandola {et~al.}(2021)Allamandola, Boersma, Lee, Bregman, \&
  Temi}]{Allamandola21}
Allamandola, L.~J., Boersma, C., Lee, T.~J., Bregman, J.~D., \& Temi, P. 2021,
  Astrophys. J., 917, L35

\bibitem[{Allamandola {et~al.}(1999)Allamandola, Hudgins, \&
  Sandford}]{Allamandola99}
Allamandola, L.~J., Hudgins, D.~M., \& Sandford, S.~A. 1999, Astrophys. J.,
  511, L115

\bibitem[{Basire {et~al.}(2011)Basire, Parneix, Pino, Br\'{e}chignac, \&
  Calvo}]{Basire11}
Basire, M., Parneix, P., Pino, T., Br\'{e}chignac, P., \& Calvo, F. 2011, in
  PAHs and the Universe, ed. C.~Joblin \& A.~Tielens, Vol.~46 (EAS Publications
  Series), 95--101

\bibitem[{Bernard {et~al.}(2011)Bernard, Seidl, Kohl, Liedl, Mayer, \'{O}scar
  G\'{a}lvez, Grothe, \& Loerting}]{Bernard11}
Bernard, J., Seidl, M., Kohl, I., {et~al.} 2011, Angew. Chem. Int. Ed., 50,
  1939

\bibitem[{Boersma {et~al.}(2011)Boersma, {Bauschlicher, Jr.}, Ricca, Mattioda,
  Peeters, Tielens, \& Allamandola}]{Boersma11}
Boersma, C., {Bauschlicher, Jr.}, C.~W., Ricca, A., {et~al.} 2011, Astrophys.
  J., 729, 64

\bibitem[{Cook \& Saykally(1998)}]{Cook98}
Cook, D.~J., \& Saykally, R.~J. 1998, Astrophys. J., 493, 793

\bibitem[{Crawford \& {Schaefer III}(2000)}]{ccreview}
Crawford, T.~D., \& {Schaefer III}, H.~F. 2000, in Reviews in Computational
  Chemistry, ed. K.~B. Lipkowitz \& D.~B. Boyd, Vol.~14 (New York: Wiley),
  33--136

\bibitem[{Davis {et~al.}(2022)Davis, Garrett, \& Fortenberry}]{Davis22}
Davis, M.~C., Garrett, N.~R., \& Fortenberry, R.~C. 2022, Phys. Chem. Chem.
  Phys., 24, 18552

\bibitem[{Douglas \& Kroll(1974)}]{Douglas74}
Douglas, M., \& Kroll, N.~M. 1974, Ann. Phys., 82, 89

\bibitem[{Dunning(1989)}]{Dunning89}
Dunning, T.~H. 1989, J. Chem. Phys., 90, 1007

\bibitem[{Finney {et~al.}(2016)Finney, Fortenberry, Francisco, \&
  Peterson}]{Finney16}
Finney, B., Fortenberry, R.~C., Francisco, J.~S., \& Peterson, K.~A. 2016, J.
  Chem. Phys., 145, 124311

\bibitem[{Fortenberry \& Lee(2019)}]{Fortenberry19QFF}
Fortenberry, R.~C., \& Lee, T.~J. 2019, Ann. Rep. Comput. Chem., 15, 173

\bibitem[{Fortenberry \& Lee(2022)}]{Fortenberry22}
---. 2022, in Vibrational Dynamics of Molecules, ed. J.~M. Bowman (Singapore:
  World Scientific), 235--295

\bibitem[{Frisch {et~al.}(2016)Frisch, Trucks, Schlegel, Scuseria, Robb,
  Cheeseman, Scalmani, Barone, Petersson, Nakatsuji, Li, Caricato, Marenich,
  Bloino, Janesko, Gomperts, Mennucci, Hratchian, Ortiz, Izmaylov, Sonnenberg,
  Williams-Young, Ding, Lipparini, Egidi, Goings, Peng, Petrone, Henderson,
  Ranasinghe, Zakrzewski, Gao, Rega, Zheng, Liang, Hada, Ehara, Toyota, Fukuda,
  Hasegawa, Ishida, Nakajima, Honda, Kitao, Nakai, Vreven, Throssell,
  Montgomery, Peralta, Ogliaro, Bearpark, Heyd, Brothers, Kudin, Staroverov,
  Keith, Kobayashi, Normand, Raghavachari, Rendell, Burant, Iyengar, Tomasi,
  Cossi, Millam, Klene, Adamo, Cammi, Ochterski, Martin, Morokuma, Farkas,
  Foresman, \& Fox}]{g16}
Frisch, M.~J., Trucks, G.~W., Schlegel, H.~B., {et~al.} 2016, Gaussian~16
  {R}evision {C}.01, gaussian Inc. Wallingford CT

\bibitem[{Gaw {et~al.}(1991)Gaw, Willets, Green, \& Handy}]{spectro91}
Gaw, J.~F., Willets, A., Green, W.~H., \& Handy, N.~C. 1991, in Advances in
  Molecular Vibrations and Collision Dynamics, ed. J.~M. Bowman \& M.~A. Ratner
  (Greenwich, Connecticut: JAI Press, Inc.), 170--185

\bibitem[{Gy\"orffy \& Werner(2018)}]{Gyorffy18}
Gy\"orffy, W., \& Werner, H.-J. 2018, J. Chem. Phys., 148, 114104

\bibitem[{Haney {et~al.}(2023)Haney, Westbrook, Santaloci, \&
  Fortenberry}]{Haney23}
Haney, O.~G., Westbrook, B.~R., Santaloci, T.~J., \& Fortenberry, R.~C. 2023,
  J. Phys. Chem. A, 127, 489

\bibitem[{Hill \& Peterson(2010)}]{Hill10}
Hill, J.~G., \& Peterson, K.~A. 2010, Phys. Chem. Chem. Phys., 12, 10460

\bibitem[{Huber {et~al.}(2012)Huber, Dalnodar, Kausch, Kimeswenger, \&
  Probst}]{Huber12}
Huber, S.~E., Dalnodar, S., Kausch, W., Kimeswenger, S., \& Probst, M. 2012,
  AIP Adv., 2, 032180

\bibitem[{Ioppolo {et~al.}(2021)Ioppolo, Kanuchov\'{a}, James, Dawes, Ryabov,
  Dezalay, Jones, Hoffmann, Mason, \& Strazzulla}]{Ioppolo21}
Ioppolo, S., Kanuchov\'{a}, Z., James, R.~L., {et~al.} 2021, Astron.
  Astrophys., 646, A172

\bibitem[{Jian {et~al.}(2021)Jian, Yang, \& Chu}]{Jian21}
Jian, H.-Y., Yang, C.-T., \& Chu, L.-K. 2021, Phys. Chem. Chem. Phys., 23,
  14699

\bibitem[{Jones {et~al.}(2014)Jones, Kaiser, \& Strazzulla}]{Jones14}
Jones, B.~M., Kaiser, R.~I., \& Strazzulla, G. 2014, Astrophys. J., 788, 170

\bibitem[{Kendall {et~al.}(1992)Kendall, Dunning, \& Harrison}]{aug-cc-pVXZ}
Kendall, R.~A., Dunning, T.~H., \& Harrison, R.~J. 1992, J. Chem. Phys., 96,
  6796

\bibitem[{Knizia {et~al.}(2009)Knizia, Adler, \& Werner}]{Knizia09}
Knizia, G., Adler, T.~B., \& Werner, H.-J. 2009, J. Chem. Phys., 130, 054104

\bibitem[{Kumar {et~al.}(2007)Kumar, Kalinichev, \& Kirkpatrick}]{Kumar07}
Kumar, P.~P., Kalinichev, A.~G., \& Kirkpatrick, R.~J. 2007, J. Chem. Phys.,
  126, 204315

\bibitem[{Mackie {et~al.}(2022)Mackie, Candian, Lee, \& Tielens}]{Mackie22}
Mackie, C.~J., Candian, A., Lee, T.~J., \& Tielens, A. G. G.~M. 2022, J. Phys.
  Chem. A, 126, 3198

\bibitem[{Mackie {et~al.}(2018{\natexlab{a}})Mackie, Chen, Candian, Lee, \&
  Tielens}]{Mackie18Cascade}
Mackie, C.~J., Chen, T., Candian, A., Lee, T.~J., \& Tielens, A. G. G.~M.
  2018{\natexlab{a}}, J. Chem. Phys., 149, 134302

\bibitem[{Mackie {et~al.}(2018{\natexlab{b}})Mackie, Candian, Huang, Maltseva,
  Petrignani, Oomens, Buma, Lee, \& Tielens}]{Mackie18}
Mackie, C.~J., Candian, A., Huang, X., {et~al.} 2018{\natexlab{b}}, Phys. Chem.
  Chem. Phys., 20, 1189

\bibitem[{Martin {et~al.}(1998)Martin, Lee, \& Taylor}]{Martin98}
Martin, J. M.~L., Lee, T.~J., \& Taylor, P.~R. 1998, J. Chem. Phys., 108, 676

\bibitem[{Mills(1972)}]{Mills72}
Mills, I.~M. 1972, in Molecular Spectroscopy - Modern Research, ed. K.~N. Rao
  \& C.~W. Mathews (New York: Academic Press), 115--140

\bibitem[{M{\o}ller \& Plesset(1934)}]{MP2}
M{\o}ller, C., \& Plesset, M.~S. 1934, Phys. Rev., 46, 618

\bibitem[{Mori {et~al.}(2009)Mori, Suma, Sumiyoshi, \& Endo}]{Mori09}
Mori, T., Suma, K., Sumiyoshi, Y., \& Endo, Y. 2009, J. Chem. Phys., 130,
  204308

\bibitem[{Mori {et~al.}(2011)Mori, Suma, Sumiyoshi, \& Endo}]{Mori11}
---. 2011, J. Chem. Phys., 134, 044319

\bibitem[{Noble {et~al.}(2011)Noble, Dulieu, Congiu, \& Fraser}]{Noble11}
Noble, J.~A., Dulieu, F., Congiu, E., \& Fraser, H.~J. 2011, Astrophys. J.,
  735, 121

\bibitem[{Papousek \& Aliev(1982)}]{Papousek82}
Papousek, D., \& Aliev, M.~R. 1982, Molecular Vibration-Rotation Spectra
  (Amsterdam: Elsevier)

\bibitem[{Pech {et~al.}(2002)Pech, Joblin, \& Boissel}]{Pech02}
Pech, C., Joblin, C., \& Boissel, P. 2002, Astron. Astrophys., 388, 639

\bibitem[{Peeters {et~al.}(2004)Peeters, Allamandola, Hudgins, Hony, \&
  Tielens}]{Peeters04}
Peeters, E., Allamandola, L.~J., Hudgins, D.~M., Hony, S., \& Tielens, A. G.
  G.~M. 2004, in Astrophysics of Dust, ASP Conference Series, ed. A.~N. Witt,
  G.~C. Clayton, \& B.~T. Draine, Vol. 309 (San Francisco, CA: Astronomical
  Society of the Pacific), 141--162

\bibitem[{Peeters {et~al.}(2010)Peeters, Hudson, Moore, \& Lewis}]{Peeters2010}
Peeters, Z., Hudson, R., Moore, M., \& Lewis, A. 2010, Icarus, 210, 480

\bibitem[{Peterson {et~al.}(2008)Peterson, Adler, \& Werner}]{Peterson08}
Peterson, K.~A., Adler, T.~B., \& Werner, H.-J. 2008, J. Chem. Phys., 128,
  084102

\bibitem[{Peterson \& Dunning(1995)}]{cc-pVXZ}
Peterson, K.~A., \& Dunning, T.~H. 1995, J. Chem. Phys., 102, 2032

\bibitem[{Peterson \& Dunning(2002)}]{Peterson2002}
---. 2002, J. Chem. Phys., 117, 10548

\bibitem[{Raghavachari {et~al.}(1989)Raghavachari, Trucks, Pople, \&
  Head-Gordon}]{Rag89}
Raghavachari, K., Trucks, G.~W., Pople, J.~A., \& Head-Gordon, M. 1989, Chem.
  Phys. Lett., 157, 479

\bibitem[{Reddy {et~al.}(2012)Reddy, Kulkarni, \& Balasubramanian}]{Reddy12}
Reddy, S.~K., Kulkarni, C.~H., \& Balasubramanian, S. 2012, J. Phys. Chem. A,
  116, 1638

\bibitem[{Sagiv {et~al.}(2018)Sagiv, Hirshberg, \& Gerber}]{Sagiv18}
Sagiv, L., Hirshberg, B., \& Gerber, R.~B. 2018, Chem. Phys., 514, 44

\bibitem[{Sandford {et~al.}(2020)Sandford, Nuevo, Bera, \& Lee}]{Sandford20}
Sandford, S.~A., Nuevo, M., Bera, P.~P., \& Lee, T.~J. 2020, Chem. Rev., 120,
  4616

\bibitem[{Sanz-Novo {et~al.}(2023)Sanz-Novo, Rivilla, Jim\'{e}nez-Serra,
  Mart\'{i}n-Pintado, Colzi, Zeng, Meg\'{i}as, \'{A}lvaro L\'{o}pez-Gallifa,
  Mart\'{i}nez-Henares, Massalkhi, Tercero, de~Vicente, Mart\'{i}n, Andr\'{e}s,
  \& Requena-Torres}]{Sanz23}
Sanz-Novo, M., Rivilla, V.~M., Jim\'{e}nez-Serra, I., {et~al.} 2023, Astrophys.
  J., \emph{accepted}

\bibitem[{Shavitt \& Bartlett(2009)}]{Shavitt09}
Shavitt, I., \& Bartlett, R.~J. 2009, Many-Body Methods in Chemistry and
  Physics: MBPT and Coupled-Cluster Theory (Cambridge: Cambridge University
  Press)

\bibitem[{Strazzulla {et~al.}(1996)Strazzulla, Brucato, \&
  Palumbo}]{Strazzulla96}
Strazzulla, G., Brucato, J.~R., \& Palumbo, M.~E. 1996, Plan. Space Sci., 44,
  1447

\bibitem[{Tielens(2008)}]{Tielens08}
Tielens, A. G. G.~M. 2008, Annu. Rev. Astron. Astrophys., 46, 289

\bibitem[{Wallace \& Fortenberry(2021)}]{Wallace21carbonic}
Wallace, A.~M., \& Fortenberry, R.~C. 2021, J. Phys. Chem. A, 125, 4589

\bibitem[{Wallace \& Fortenberry(2022)}]{Wallace22}
---. 2022, J. Phys. Chem. A, 126, 3739

\bibitem[{Wang \& B\"{u}rgi(2021)}]{Wang21}
Wang, X., \& B\"{u}rgi, T. 2021, Angew. Chem. Int. Ed., 60, 7860

\bibitem[{Watrous {et~al.}(2021)Watrous, Westbrook, \& Fortenberry}]{Watrous21}
Watrous, A.~G., Westbrook, B.~R., \& Fortenberry, R.~C. 2021, J. Phys. Chem. A,
  125, 10532

\bibitem[{Watson(1977)}]{Watson77}
Watson, J. K.~G. 1977, in Vibrational Spectra and Structure, ed. J.~R. During
  (Amsterdam: Elsevier), 1--89

\bibitem[{Werner {et~al.}(2012)Werner, Knowles, Knizia, Manby, \&
  Sch{\"u}tz}]{MOLPRO-WIREs}
Werner, H.-J., Knowles, P.~J., Knizia, G., Manby, F.~R., \& Sch{\"u}tz, M.
  2012, WIREs Comput. Mol. Sci., 2, 242

\bibitem[{Werner {et~al.}(2022)Werner, Knowles, Knizia, Manby, Schütz, Celani,
  Gy\"{o}rffy, Kats, Korona, Lindh, Mitrushenkov, Rauhut, Shamasundar, Adler,
  Amos, Bennie, Bernhardsson, Berning, Cooper, Deegan, Dobbyn, Eckert, Goll,
  Hampel, Hesselmann, Hetzer, Hrenar, Jansen, K\"{o}ppl, Lee, Liu, Lloyd, Ma,
  Mata, May, McNicholas, Meyer, {Miller III}, Mura, Nickla{\ss}, O'Neill,
  Palmieri, Peng, Pfl\"{u}ger, Pitzer, Reiher, Shiozaki, Stoll, Stone, Tarroni,
  Thorsteinsson, Wang, \& Welborn}]{MOLPRO22}
Werner, H.-J., Knowles, P.~J., Knizia, G., {et~al.} 2022, see
  http://www.molpro.net

\bibitem[{Westbrook \& Fortenberry(2020)}]{Westbrook20}
Westbrook, B.~R., \& Fortenberry, R.~C. 2020, J. Phys. Chem. A, 124, 3191

\bibitem[{Westbrook \& Fortenberry(2023)}]{Westbrook23}
---. 2023, J. Chem. Theory Comput., 19, 2606

\bibitem[{Yousaf \& Peterson(2008)}]{Yousaf08}
Yousaf, K.~E., \& Peterson, K.~A. 2008, J. Chem. Phys., 129, 184108

\bibitem[{Zheng \& Kaiser(2007)}]{Zheng07}
Zheng, W., \& Kaiser, R.~I. 2007, Chem. Phys. Lett., 450, 55

\bibitem[{Zhu {et~al.}(2011)Zhu, Hosokai, \& Akiyama}]{Zhu11}
Zhu, C., Hosokai, S., \& Akiyama, T. 2011, Crys. Growth Des., 11, 4166

\end{thebibliography}

\end{document}